\documentclass[10pt]{iopart}
\usepackage{graphicx}
\def \pbpb  {Pb+Pb}
\def \pp    {$pp$ }
\def\corr{\langle \Delta p_{t,1}, \Delta p_{t,2}\rangle}

\begin{document}
\title{Heavy--quark momentum correlations as  a sensitive probe of thermalization 
%\thanks{Presented at ISMD07, Berkeley, CA}%
}
\author{K. Schweda and G. Tsiledakis}
\address{Physikalisches Institut, Universit\"at Heidelberg, D-69120 Heidelberg, Germany}

\begin{abstract}
In high-energy nuclear collisions 
the degree of thermalization at the partonic level is a key issue. 
Due to their large mass,
heavy--quarks and their participation in the collective flow of the QCD medium
constitute a powerful tool to probe thermalization.
%In strong interactions, heavy--quarks are always created together with their anti--quark and are thus correlated.
We propose measuring azimuthal correlations of heavy-quark hadrons and products from their semi-leptonic decay.
Modifications or even the complete absence of initially, e.g. in $pp$ collisions, existing azimuthal correlations in \pbpb\ 
collisions might indicate thermalization at the partonic level. We present studies with PYTHIA for
$pp$ collisions at the top LHC energy using the two-particle transverse momentum correlator 
${\langle\overline{\Delta}p_{t,1}\overline{\Delta}p_{t,2}\rangle}$
 as a sensitive measure of  azimuthal
correlations. 
%The influence due to collective flow present in collision of heavy nuclei is estimated.
\end{abstract}
%\PACS{PACS numbers come here}
 %

 \section{Introduction}
High-energy nuclear collisions offer the unique opportunity to probe highly
excited nuclear matter in the laboratory. At sufficiently high
temperature and/or energy density hadrons dissolve 
and quarks and gluons carrying color charge
are the relevant degrees of freedom, commonly called
a Quark Gluon Plasma (QGP).
An essential difference between collisions of elementary particles on the one hand
and heavy nuclei on the other hand is the development of collectivity in the
latter. Collective flow of hadrons, especially the multi--strange hadrons $\phi$
and $\Omega$, has been experimentally observed at RHIC~  \cite{strange} suggesting that collectivity
dominantly develops in the early partonic stage, i.e. among quarks and gluons. Presently, the degree of
thermalization among partons remains a crucial issue.
\begin{figure}[htb]
\begin{minipage}{0.50\textwidth} 
%\begin{center}
%\centering
%\includegraphics[width=0.96\textwidth]{../dNdphi.pdf}
\includegraphics[width=0.99\textwidth]{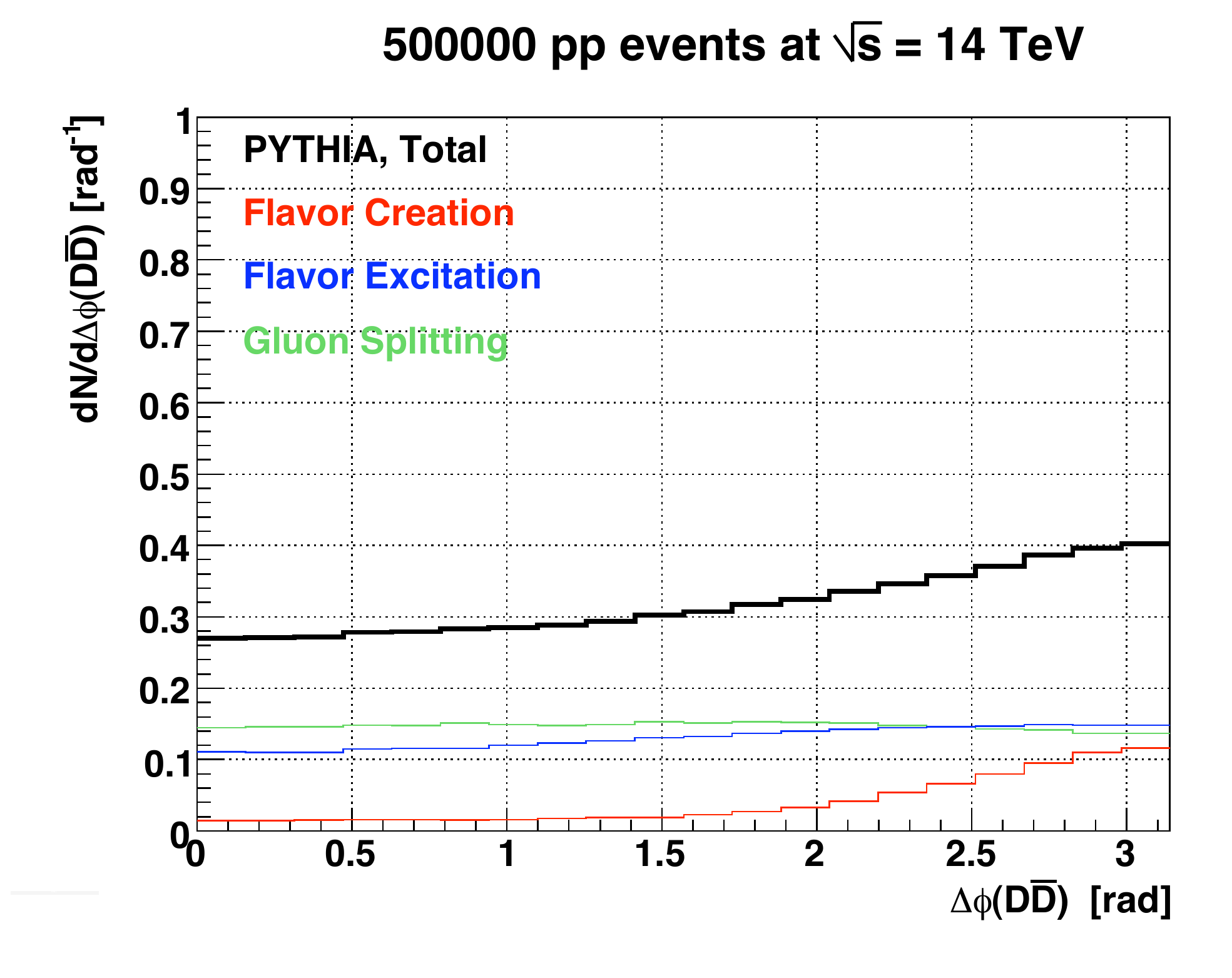}
\label{dndphi}
%\end{center}
\end{minipage}
\begin{minipage}{0.49\textwidth} 
\caption{(Color online) Distribution of $D\overline{D}$ pairs 
as  a function of relative azimuth~$\Delta\phi$
from $pp$ collisions at $\sqrt{s}$ = 14 TeV as calculated by PYTHIA (v. 6.406)
integrated over rapidity from flavor creation~(red line) , flavor excitation~(blue line), gluon
splitting~(green line), and their sum~(black line). }
\end{minipage} 
\end{figure}

Heavy--quark hadrons and their observables are of particular interest when addressing
thermalization~\cite{thermalization}. In a QGP, where chiral symmetry might be partially restored 
and consequently light--quarks $(up, down, strange)$ obtain their small bare masses, 
heavy--quarks remain massive~\cite{Zhu}. 
Heavy--quarks are mostly created in the early stage of the collision, since their mass is much larger
than the maximum possible temperature of the QCD medium created while annihilation 
in the QCD medium is negligible~\cite{pbm_2007}. 
Thus, heavy--quarks probe the entire history of a high-energy nuclear collision.
Heavy--quarks participate in collective motion provided their
interactions at the partonic level occur frequently. Thus, collective motion
of heavy-quark hadrons is a powerful tool when addressing early thermalization of
light quarks in high-energy nuclear collisions.

\subsection{$D\overline{D}$ angular correlations}
In strong interactions
heavy--quarks are always created together with their anti--quark and are thus correlated.
In collisions of elementary particles, these correlations survive the fragmentation
process into hadrons to a large extent~\cite{Zhu} and are thus observable e.g. in the angular distributions 
of pairs of $D$-- and $\overline{D}$--mesons ~\cite{review}. 

In high--energy collisons of heavy nuclei, frequent interactions among partons (quarks and gluons) of the QCD 
medium and heavy--quarks
would lead to a weakening and finally complete vanishing of these correlations.
On the other hand, hadronic interactions at the late stage are insufficient
to alter azimuthal correlation of 
$D\overline{D}$--pairs \cite{Zhu}. 
It is frequent interactions
distributing and randomizing the available (kinetic) energy and finally driving the system, i.e. light--quarks and gluons,
to local thermal equilibrium. Thus, the decreasing magnitude when compared to \pp collisions of heavy--quark correlations in  high--energy collisons of heavy nuclei would indicate early thermalization among partons.

We employed the Monte Carlo event generator PYTHIA \cite{pythia} which reproduces 
experimentally observed correlations of $D$ mesons at fixed target energies~\cite{review}.
Parameters in PYTHIA were tuned  to reproduce the NLO 
predictions~\cite{pythia2,Dainese}.
Our results on the yield of $D\overline{D}$--pairs versus their relative azimuth
for \pp collisions at the top LHC energy of $\sqrt{s}$ = 14 TeV
are shown in  Fig.~\ref{dndphi}.
Our calculations at leading order (LO)  contain flavor
creation processes ($q\overline{q}\rightarrow Q\overline{Q}$, $gg\rightarrow Q\overline{Q}$)
and lead to an enhancement around $\Delta\phi \approx 180^{\rm o}$ , i.e. back-to-back .
However, next-to-leading order (NLO) contributions such as flavor excitation
($qQ\rightarrow qQ, gQ\rightarrow gQ$) and gluon splitting ($g\rightarrow Q\overline{Q}$) 
become dominant at LHC energies and do not show pronounced 
correlations in the representation chosen in Fig.~\ref{dndphi} leading to a rather weak dependence on relative azimuth.
By choosing various momentum cuts, a somewhat stronger correlation was extracted~\cite{Zhu2}. 

In this paper, we introduce
the two-particle transverse momentum correlator as a sensitive measure of heavy-quark correlations. 
This method has the following advantages: \newline
(i) the correlator is sensitive to non-statistical fluctuations, thus carving out any physical correlations and
eliminating the need for various momentum-cuts in the analysis. This is essential when comparing results from different
collision systems, e.g. \pp and \pbpb. \newline
(ii) in case of physically uncorrelated candidate--pairs (e.g. background), the extracted value for the correlator is null,
thus providing a reliable baseline. 

\section{Employing the two-particle transverse momentum correlator}
The occurence of non-statistical fluctuations of the event-by-event mean 
transverse momentum $M_{pt}$ goes along 
with correlations among the transverse momenta of particle pairs. 
Such correlations were successfully extracted from experimental data
employing the two-particle transverse momentum correlator~\cite{pipj,ceres-pt}.
For $D$-- and $\overline{D}$--mesons, this leads to
\begin{equation}
\corr^{(D\overline{D})}=
\frac{\sum_{k=1}^{n_{\rm ev}}  C_{k}}  
{\sum_{k=1}^{n_{\rm ev}}N_k^{\rm pairs}}
%\sum_{k=1}^{n_{\rm ev}}
%\sum_{i=1}^{N_k}\sum_{j=1}^{N_k}(p_{ti}-\overline{p_t})
%(p_{tj}-\overline{p_t}) .
\end{equation}
where $C_{k}$ is the $p_{t}$ convariance:
\begin{equation}
C_{k}=
%\sum_{k=1}^{n_{\rm ev}}
\sum_{i=1}^{N_k}\sum_{j=1}^{N_k}(p_{ti}-\overline{p_t}^{(D)})
(p_{tj}-\overline{p_t}^{(\overline{D})}) 
\end{equation}
where $p_{ti}$ and $p_{tj}$ are the transverse momentum $p_{t}$ for $i^{th}$ and $j^{th}$ 
particle of an event of $D$-- and $\overline{D}$--mesons respectively,
$\overline{p_t}$ is the inclusive mean transverse momentum 
averaged over all particles of all events of $D$ and $\overline{D}$, 
 $\sum_{k=1}^{n_{\rm ev}}N_k^{\rm pairs}$ the total number of 
$D\overline{D}$ pairs and $n_{ev}$ the total number of \pp collisions.
%\begin{equation}
%\corr=
%\frac{1} 
%{\sum_{k=1}^{n_{\rm ev}}N_k^{\rm pairs}}.
%\sum_{k=1}^{n_{\rm ev}}
%\sum_{i=1}^{N_k}\sum_{j=1}^{N_k}(p_{ti}-\overline{p_t})
%(p_{tj}-\overline{p_t}) .
%\end{equation}
%
\begin{figure}[!th]
 \begin{center}
		\includegraphics[width=0.7\textwidth]{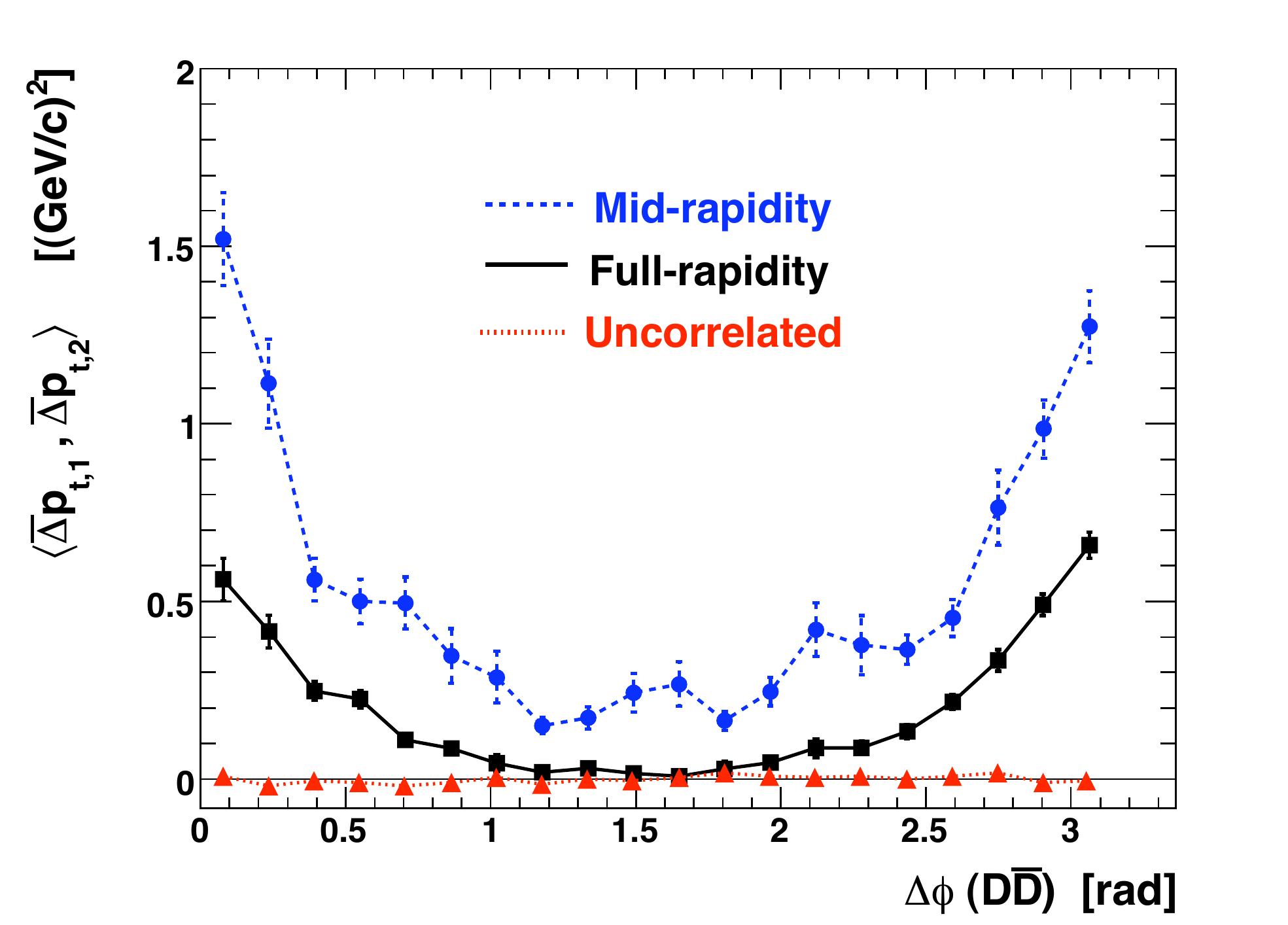}
\end{center}
\caption{(Color online) Distribution of  the momentum correlator $\corr$ of $D\overline{D}$ pairs  
as a function of relative azimuth $\Delta\phi$ 
at mid-rapidity~(circles), integrated over rapidity~(squares) and for background using the mixed event method (triangles) 
from $pp$ collisions at $\sqrt{s}$ = 14 TeV as calculated by PYTHIA (v. 6.406). The lines are drawn to guide the eye.} 
\label{Ckpipj}
\end{figure}

We studied the scale dependence of $p_{t}$ correlations in
azimuthal space by calculating the correlator in bins of the relative azimuthal separation 
$\Delta\phi$ of particle pairs.
For the case of independent particle emission from a single parent distribution, the correlator $\corr$
vanishes. 

The $D\overline{D}$ momentum correlator $\corr$ as a function of relative
azimuth $\Delta\phi$ is shown in Fig.~\ref{Ckpipj} for \pp collisions at $\sqrt{s}$ =~14~TeV exhibiting a rich structure.
We observe an enhancement at small azimuth from gluon
splitting processes, while flavor creation of $c\overline{c}$--quark pairs leads to a pronounced distribution at backward angels.
Flavor excitation processes involve a larger number of gluons leading to a rather flat distribution
At mid-rapidity, the correlations are even stronger, reflecting the harder $p_{t}$ spectrum of $D\overline{D}$ meson-pairs
when compared to the full rapidity range.
Integrating the correlator over all azimuth,  we get
$\corr = 0.199\pm0.006$~GeV$^2$/$c^2$  which corresponds to the normalized dynamical 
fluctuation $\Sigma_{pt}$ \cite{spt} of $\sim30\%$ in $\overline{p_t}$. 
This large value implies a strong
correlation when compared to 
$\sim1\%$ observed for unidentified charged particles in central collisions at SPS and RHIC \cite{spt,spt1,spt2}. 
To mimic combinatorial background which is always present in the experiment, we applied the correlator 
to  $D$-- and $\overline{D}$-- mesons  from different \pp collisions, which are physically uncorrelated. 
This results in a value of $\corr$ consistent with zero. Therefore the correlator
allows for a clear distinction between the case were correlations are present (different from zero) or absent (equal to zero)
in contrast to the method described in~\cite{Zhu}.
%
%Our tests show that a realistic amount of elliptic flow does not change these
%correlations. Concerning the radial flow contribution, it is assumed that the 
%expansion produces an additional momentum $p_{t,f}=\gamma m \beta$, where $\gamma$ is the Lorentz
%factor, $\beta$ is the profile velocity and $m$ is the mass of the $D$ meson.
%By adding this radial flow component \cite{flow} vectorially to the momentum vector produced by PYTHIA, we 
%evaluate the $\corr$ as a function of $\Delta\phi$. 
\begin{figure}
\begin{minipage}{0.49\textwidth} 
%\begin{center}
\centering
\vspace{0.cm} 
\includegraphics[width=0.999\textwidth]{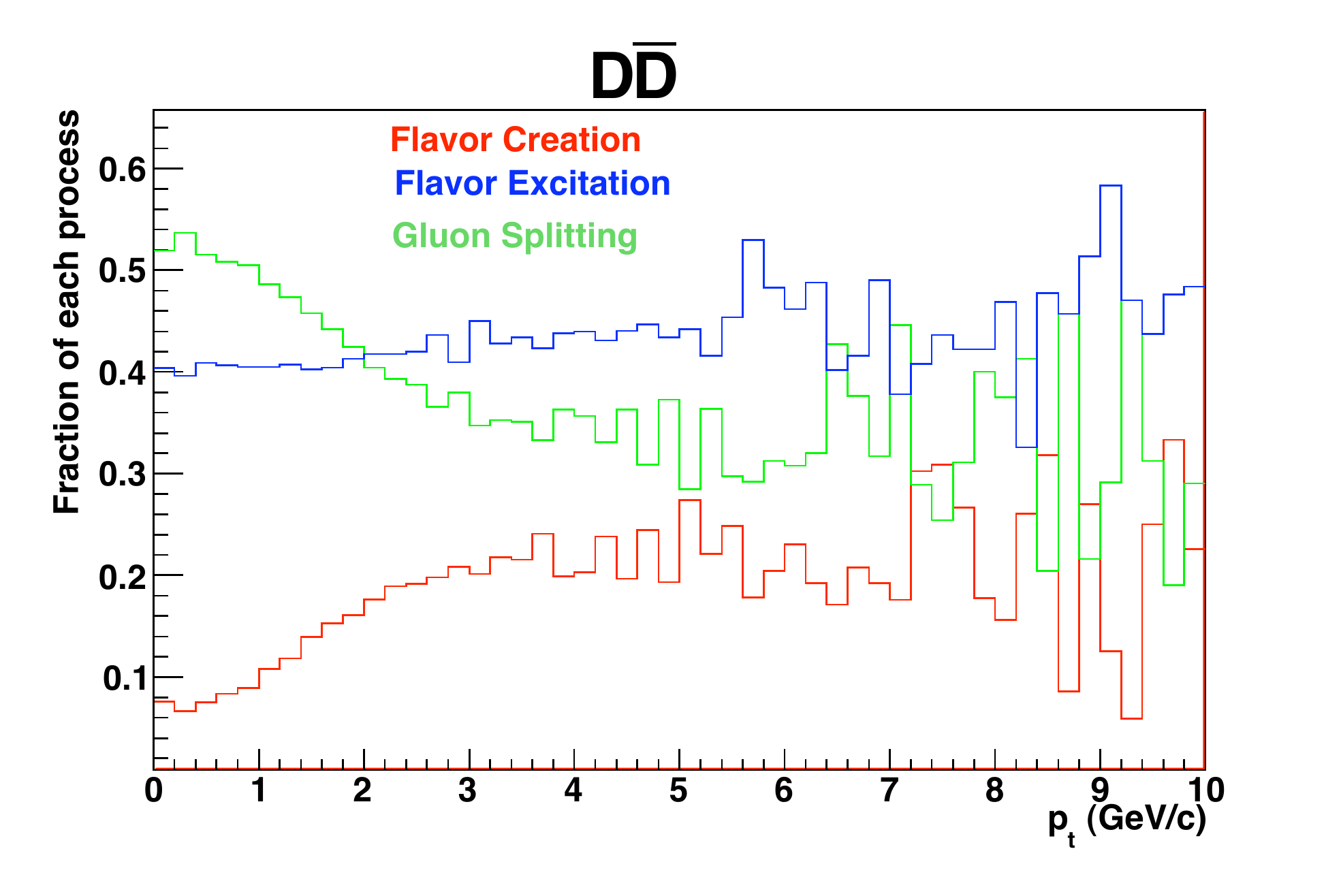}
%\end{center}
\end{minipage}
\begin{minipage}{0.49\textwidth}
\centering
\includegraphics[width=0.999\textwidth]{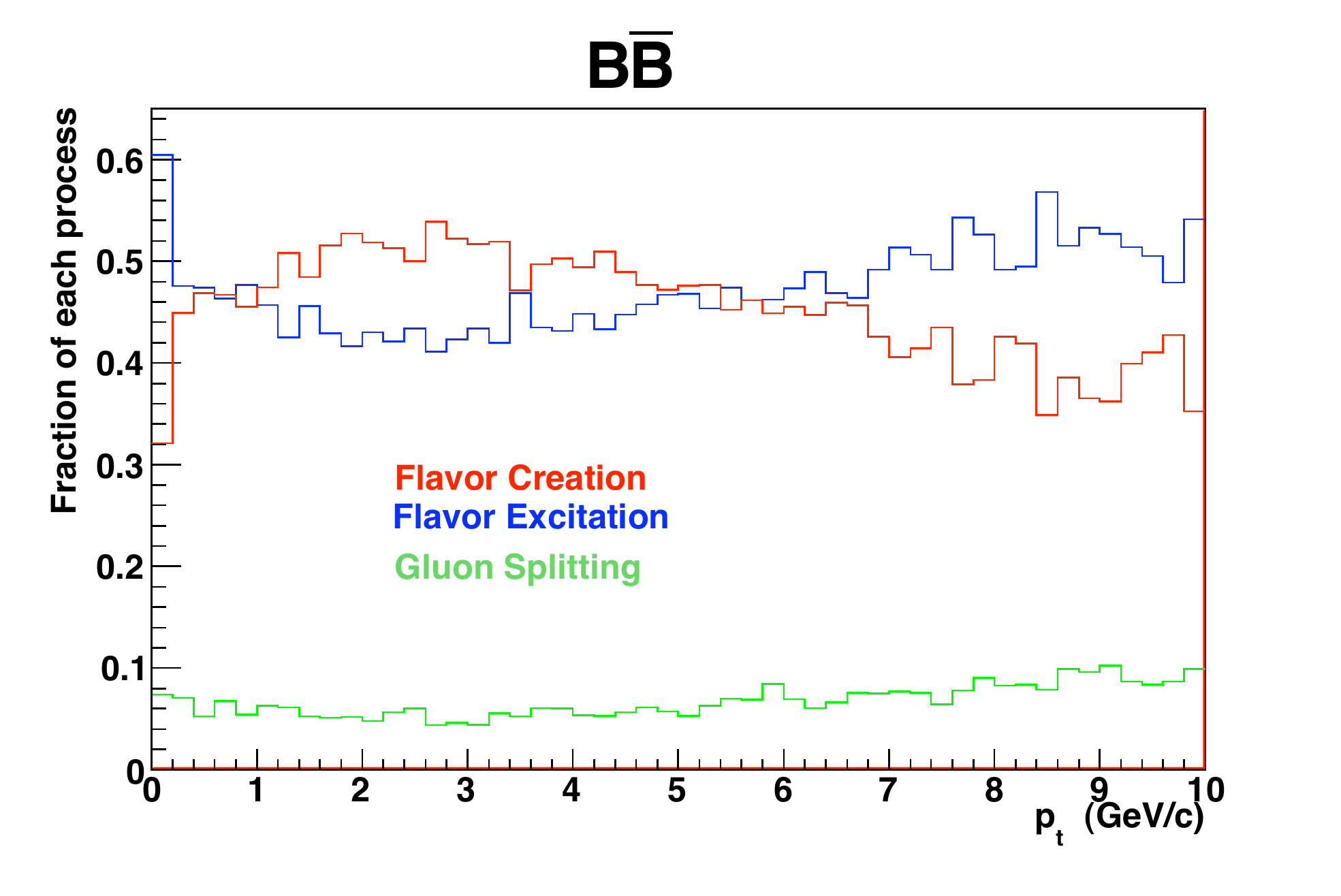}
\end{minipage} 
\centering
 \caption{(Color online) 
 Relative contribution to the production of $D$-- and $\overline{D}$--mesons~(left) 
 and $B$-- and $\overline{B}$--mesons~(right) 
 as a function of 
 transverse momentum for the processes of  
 flavor creation~(red line) , flavor excitation~(blue line) and gluon
 splitting~(green line) from $p-p$ collisions at $\sqrt{s}$ = 14 TeV as calculated by PYTHIA (v. 6.406). 
}
\label{radial}
\end{figure}
%As it is shown in Fig.~\ref{radial}, strong radial flow will further increase
%the same side momentum correlations of $D\overline{D}$ pairs and might lead to strong 
%anti-correlations at large angles.
The different QCD processes contributing to heavy--quark productions have strong energy dependences.
Figure~\ref{radial} shows the relative contribution to the production of $D$-- and $\overline{D}$--mesons~(left) 
 and $B$-- and $\overline{B}$--mesons~(right) as a function of transverse momentum.
The production of $D$--mesons is dominated by gluon splitting and flavor excitation processes while the contribution
from flavor creation is about 10\% at low momentum and increases up to 20\% at larger momentum. At RHIC energies, the 
contribution from gluon splitting was estimated by the STAR experiment to be less than 10\%~\cite{mischke}. 
On the other hand, the production of $B$--mesons is dominated by flavor creation and flavor excitation with a small contribution
from gluon splitting below 10\% and an overall weak dependence on transverse momentum. 

As shown above, the initial correlations of $c\overline{c}$--quark pairs survive the fragmentation process to a large extent. 
However experimentally, full kinematic reconstruction of $D$--mesons from topological decays suffer from small efficiencies resulting in low statistics,
especially when pairs of  $D$--mesons are considered where the reconstruction efficiency enters quadratically.
To circumvent this fact, we considered electrons (positrons) from semi-leptonic decays of $D$-- and $B$--mesons with a
branching ratio to electrons of 10\% and 11\%, respectively.

\begin{figure}
\begin{center}
	\includegraphics[width=0.6\textwidth]{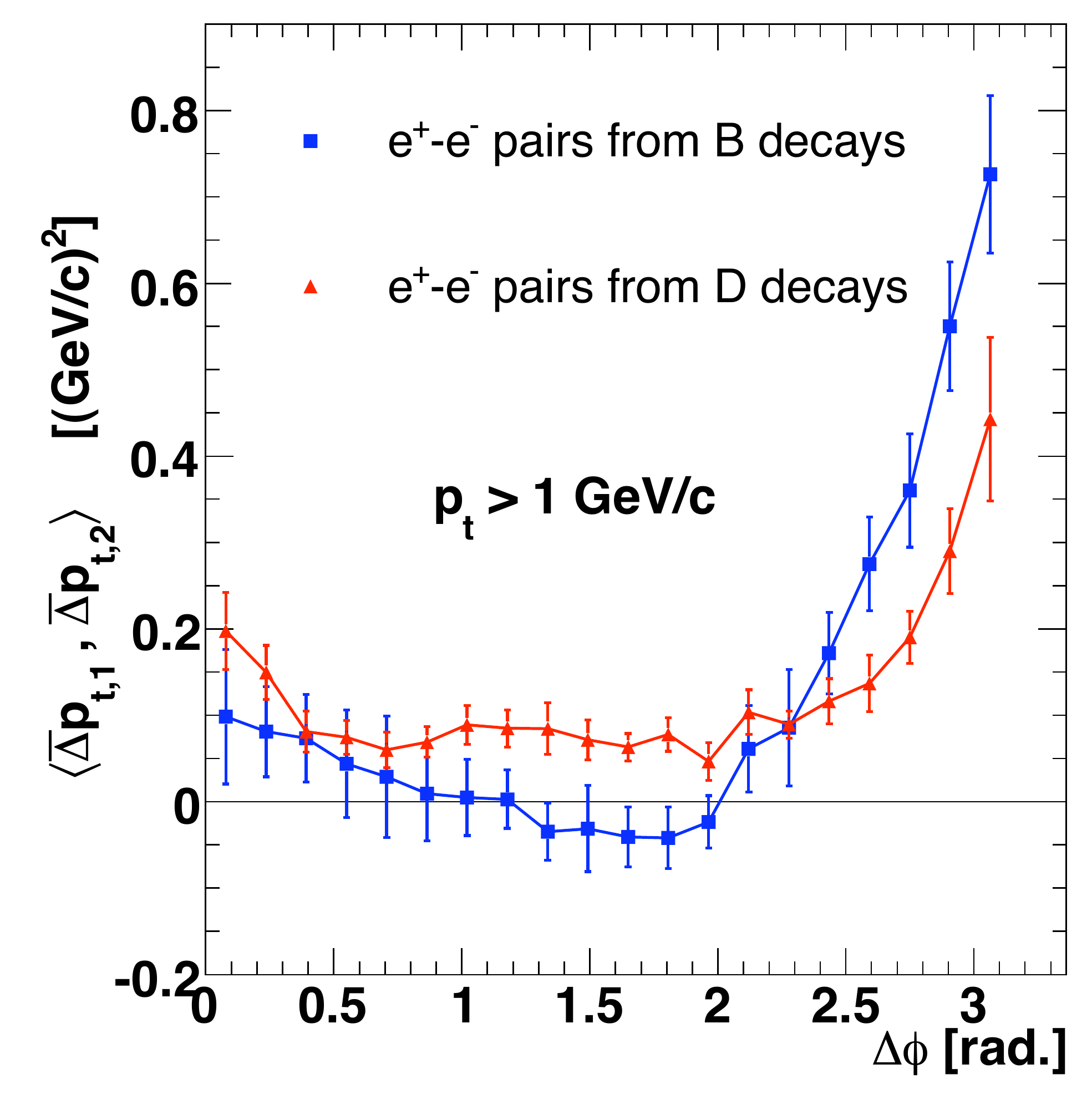}
\end{center}
\caption{(Color online) Distribution of  the momentum correlator $\corr$ of   $e^+e^-$ pairs
stemming from semi-leptonic decays of pairs of $B\overline{B}$--mesons~(squares) and 
$D\overline{D}$--mesons~(triangles)
as a function of relative azimuth $\Delta\phi$ 
at mid-rapidity 
in $pp$ collisions at $\sqrt{s}$ = 14 TeV as calculated by PYTHIA (v. 6.406). 
The lines are drawn to guide the eye.}
\label{elec}
\end{figure}

Figure~\ref{elec} shows the momentum correlator $\corr$ of $e^{+}e^{-}$--pairs from decays of $D\overline{D}$~(squares)
and $B\overline{B}$--mesons pairs~(triangles)  as a function of their relative azimuth $\Delta\phi$ 
integrated over the full rapidity range. We chose a momentum cut of $p_t > 1$GeV/$c$ allowing for clean electron identification at reasonable background and still sufficient statistics in the ALICE experiment using the Transition Radiation Detector.
We observe a correlation at forward angles due to gluon splitting and a more pronounced correlation at away-side angles
from flavor creation processes indicating that the initial correlations among heavy--quarks and their anti--quark even survive the
semi-leptonic decay into electrons (positrons) making  an experimental observation at the LHC possible. We note that the 
correlation is stronger for electrons from semi-leptonic decays of $B$--mesons compared to $D$--mesons. This is in agreement
with results from the STAR experiment at RHIC~\cite{STAR_be}.

When integrating over full azimuth for $D$--mesons, we get a value for the correlator of  $\corr =
0.007\pm0.001$~GeV$^2$/$c^2$, which corresponds to the normalized dynamical 
fluctuation $\Sigma_{pt}$ of $\sim12\%$. Hence, more than half of the correlation strength observed for fully reconstructed
$D$--meson pairs is still present in pairs of electrons and positrons from semi-leptonic decays.

\section{Conclusions and outlook}
In summary, we propose measuring correlations of heavy--quarks and their modifications in \pbpb\ collisions compared to \pp collisions
as a sensitive probe of thermalization at the early partonic stage. We applied the momentum correlator as a precise and normalized
measure. Different QCD processes contributing to heavy--quark production are identified in the azimuthal distribution of the correlator.
Our results indicate that these correlations born at the creation of pairs of heavy-quarks and their anti-quarks survive
the fragmentation in  \pp collisions into hadrons and even semi-leptonic decays into electrons to a large extent. Thus, experimental observation of these correlations in \pp and \pbpb\ collisions at the LHC seems feasible.

\section{Acknowledgment}
We would like to thank Drs.~Y.~Pachmayer and ~N.~Xu for exciting discussions.
This work has been supported by the Helmholtz Association under contract 
number VH-NG-147.
% and the Federal Ministry of Education and Research under promotional reference 06HD197D.

\vspace{3mm}
%\bibitem{karsch} 
%F. Karsch, Nucl. Phys. {\bf A698} (2002) 199c.
%Z. Fodor and S.D. Katz, J. High Energy Phys. {\bf03} (2002) 014.
%F. Karsch, Lect. Notes Phys. {\bf 583} (2002) 209.


\begin{thebibliography}{99}
\bibitem{strange} 
%K.H. Ackermann {\it et al.} (STAR collaboration), Phys. Rev. Lett. {\bf 86}, 402 (2001); 
%C. Adler {\it et al.} (STAR collaboration), Phys. Rev. Lett. {\bf 89}, 132301 (2002); 
J. Adams {\it et al.} (STAR collaboration), Phys. Rev. Lett. {\bf 95}, 122301 (2005); \\
B.I. Abelev {\it et al.} (STAR collaboration), Phys. Rev. Lett. {\bf 99}, 112301 (2007).
\bibitem{thermalization} O.~Linnyk, E.L.~Bratkovskaya, and W.~Cassing, 
Int. J. Mod. Phys.  {\bf E17}, 1367 (2008). 
\bibitem{Zhu} X. Zhu {\it et al.}, Phys. Lett. B {\bf 647}, 366 (2007); X.~Zhu, {\it these proceedings}.
%\bibitem{shuryak}
\bibitem{pbm_2007} P. Braun-Munzinger, J. Phys. G {\bf 34} S471 (2007).
\bibitem{review}C. Louren\c{c}o and H.K. W\"ohri, Phys. Rept. {\bf 433}, 127 (2006). 
\bibitem{pythia}T. Sj\"ostrand {\it et al.}, Comput. Phys. Commun {\bf 135}, 238 (2001). 
\bibitem{pythia2}E. Norrbin and T. Sj\"ostrand, Eur. Phys. J. C {\bf 17}, 137 (2000). 
\bibitem{Dainese}N. Carrer and A. Dainese (ALICE Collaboration), arXiv:hep-ph/0311225.
\bibitem{Zhu2} X. Zhu {\it et al.}, Phys. Rev. Lett. {\bf 100}, 152301 (2008).
\bibitem{pipj}J. Adams {\it et al.} (STAR collaboration), Phys. Rev. C {\bf 72}, 044902 (2005).
\bibitem{ceres-pt} D. Adamova {\it et al.} (CERES collaboration), Nucl. Phys. A {\bf 811}, 179 (2008).
\bibitem{spt}    D. Adamova {\it et al.} (CERES collaboration), Nucl. Phys. A {\bf 727}, 97 (2003).
\bibitem{spt1}H. Sako {\it et al.} (CERES collaboration), J. Phys. G {\bf 30}, S1371 (2004).
\bibitem{spt2}M. Rybczynski {\it et al.} (NA49 collaboration), J. Phys. Conf. Ser.  {\bf 5}, 74 (2005).
\bibitem{mischke} A. Mischke  {\it et al.} (STAR collaboration), J. Phys. G {\bf 35},  104117 (2008). 
\bibitem{STAR_be} X. Lin  {\it et al.} (STAR collaboration), Nucl. Phys A {\bf 783} 497 (2007).

%\bibitem{ko_2007}

%\bibitem{heinz_2007}

\end{thebibliography}
\end{document}